\documentclass[a4paper,11pt]{article}
\usepackage{graphicx}
\usepackage{amsmath}
\usepackage{amssymb}
\usepackage{afterpage}
\usepackage{verbatim}
\usepackage{psfrag}
\usepackage{color}
\usepackage{cite}
\usepackage{lettrine}
\usepackage[utf8]{inputenc}
\usepackage[T1]{fontenc}
\usepackage{authblk}
\usepackage{setspace}
\usepackage{epsfig}
\usepackage{epstopdf}
\usepackage{footmisc}
\usepackage{fmtcount}
\usepackage{stackrel}
\usepackage{float}
\usepackage{breqn}
\usepackage{enumerate}
\usepackage{subcaption}

\usepackage[ruled,vlined,linesnumbered]{algorithm2e}
\SetKwInput{KwInput}{Input}
\SetKwInput{KwOutput}{Output}
\date{}
\begin{document}

\title {Climate Monitoring using Internet of X-Things}
\author{Nasir Saeed, Tareq Y. Al-Naffouri, and  Mohamed-Slim Alouini}
\affil{Computer Electrical and Mathematical Sciences \& Engineering (CEMSE) Division, King Abdullah University of Science and Technology (KAUST), Saudi Arabia \par
e-mail: \{nasir.saeed, tareq.naffouri, slim.alouini\}@kaust.edu.sa}


\maketitle{}

\begin{abstract}
Global climate change is significantly affecting the life on planet Earth. Predicting timely changes in the climate is a big challenge and requires great attention from the scientific community. Research now suggests that using the internet of X-things  (X-IoT) helps in monitoring global climate change.
\end{abstract}

\section*{}
With the onset of climate change, climate monitoring represents one of the biggest challenges facing the world today. Global climate change directly impacts various economic, environmental, and societal aspects such as safety, food security, and energy \cite{Rineau2019}. A recent report of the united nations development program (UNDP) shows that between 1998 to 2017, climate-related disasters caused death to 1.3 million people, and left 4.4 billion injured \cite{un2019}. Besides, it also causes an average economic loss of hundreds of billions of dollars annually. In this context, it was not a surprise that climate action was added to the global sustainable development goals (SDGs) by the UNDP \cite{un2019}.
Thus, predicting the magnitude of future impacts is of paramount importance. Many variables affect global climate, such as ocean salinity, oceanic $\text{CO}_2$ emissions, land dynamics, cloud properties, atmospheric temperature, and lightening. Therefore, monitoring these variables is vital to understand changes in the global environment. 
Information communication technologies (ICTs) are beneficial for monitoring these climate variables \cite{Messer713}. In this regard, the World Meteorological Organization (WMO) and International Telecommunication Union (ITU) outline the potential of ICTs for climate monitoring \cite{Ospin2014}. At present, ICTs such as sensing instruments on satellites, sensors in oceans, and weather RADARs are the primary sources that provide climate change information. Several studies show that the use of various internet of things (IoT), including internet of underwater things (IoUT) \cite{kao2017comprehensive}, internet of underground things (IoUGT) \cite{saeed2019towards}, and internet of space things (IoST) \cite{AKYILDIZ2019134} can be of great benefit in studying these critical climate-changing variables . For simplicity, here we use the framework of the internet of X-things (X-IoT) that combines information from all these different types of IoT networks (see {\it Fig. 1\/}).
\begin{figure}
	\begin{center}
		\includegraphics[width=1\columnwidth]{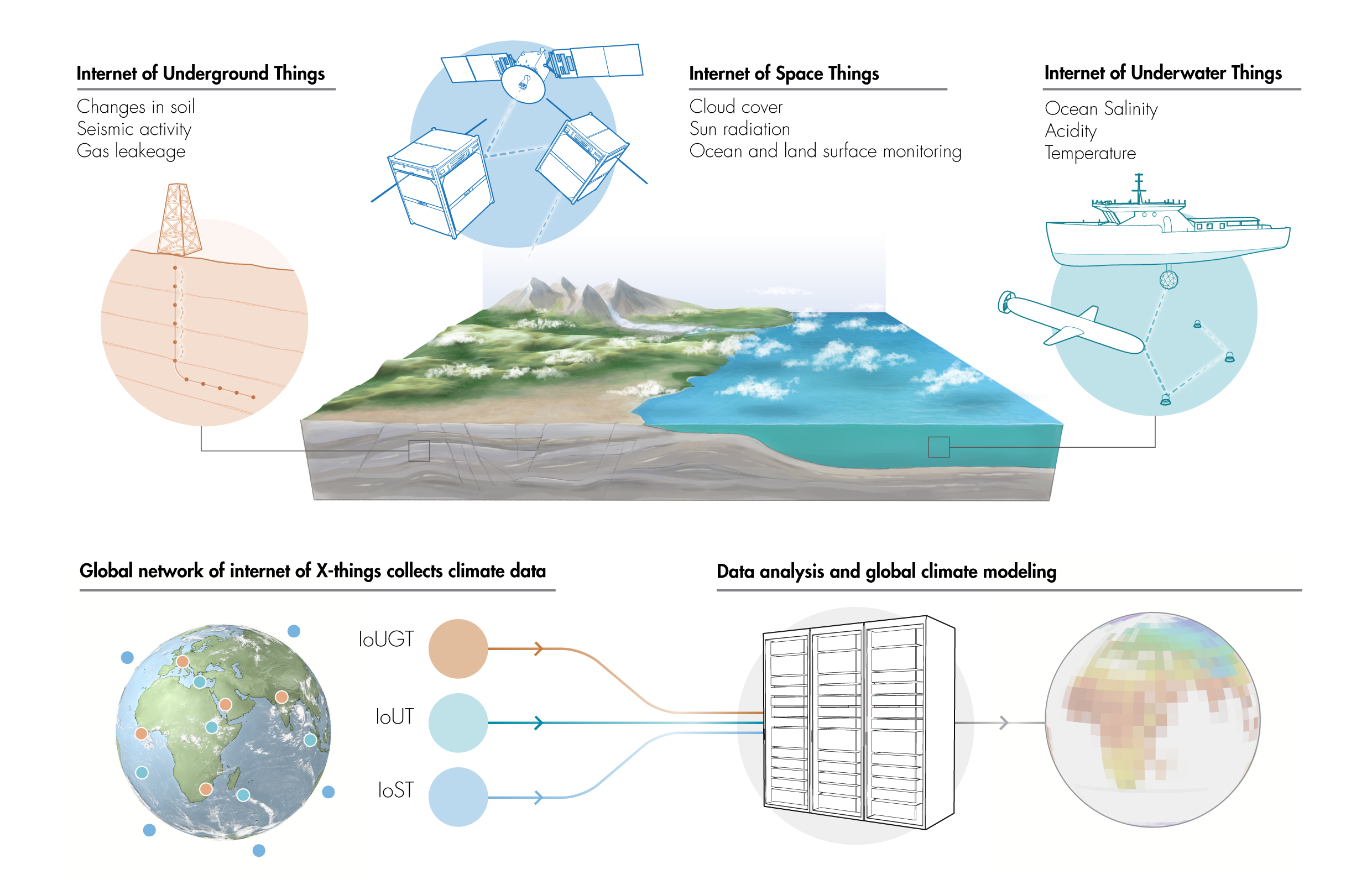}
		\caption{Climate monitoring using internet of X-things. The three different types of IoT networks collects the climate data that can help in predicting global climate change.}
	\end{center}
\end{figure}

The X-IoT framework consists of smart sensors in different radical spaces, i.e., in oceans, underground, and in space. For example, the sensors in oceans measure different variables such as ocean salinity, acidity, and temperature. Likewise, buried sensors can provide information about the underground changes in soil, seismic activities, and gas sensing. In the same way, the satellites equipped with the sensors provide information such as oceans altitude, cloud properties, amount of sun radiations, spatiotemporal knowledge of lands, and CO2 emission from
oceans. Combining all this sensing information can constitute a revolutionary advancement in climate monitoring, yet it is a challenging task because of issues with communication between networks and data analysis of such a large amount of data. Below we separately describe the different challenges facing X-IoT, from issues related to sensing and communication, localization and the type of sensing devices.

\section*{\small{Sensing and Communication}}
Sensing and communication are achieved in different ways depending on the type of environment in which the individual network exists as part of the overall X-IoT framework. For example, IoUT networks consist of underwater sensors, autonomous underwater vehicles (AUVs), surface buoys, and surface stations. The sensors are either attached to the seabed or floating in oceans. Centralized and distributed architectures are mainly used to communicate the sensing information to the outside world. In a centralized setup, AUVs or surface buoys collect the data from the underwater sensors and share it with the surface station. While in a distributed configuration, the sensors are floating and perform dive and rise operation to sense and transmit the information, respectively.

These sensors use various communication technologies for information sharing, such as acoustic, optical, and magnetic induction technologies. For instance, IoUT devices operating on acoustic waves can achieve long-range communications, albeit with a low data rate due to the low propagation speed of acoustic waves in water. On the other hand, optical waves can provide high-speed underwater communications, except only over short-range distances  \cite{SAEED2019101935}. Underwater impediments such as attenuation, scattering, and turbulence limit the range of optical waves. Alike, studies show that magnetic
induction provides high-speed and short-range underwater communications \cite{Akyildiz2015}. However, it
requires stringent alignment between the coils, which is very hard to achieve in the marine environment.
With all these different possible solutions for underwater communication, acoustic
technology is old and mature, while others are still in the academic research phase. Acoustic technology is currently the most mature technology, while the others are still in the academic research phase.

Consequently, monitoring of the subsurface environment is also critical for the climate change. For instance, geological leakage of subterranean methane and $\text{CO}_2$ is a major source of greenhouse gas (GHG) emissions. Building a network of underground sensors to measure these emissions is thus very important. Unlike terrestrial or underwater IoT networks, there are limited communication options for IoUGT because of the difficult environment. Recent studies show that magnetic induction is a promising technology for underground wireless communications \cite{ryu2018method}. The coils are attached to the subsurface sensors where the change of the magnetic field in one coil induces a current in the other and shares the information this way. Although the channel for magnetic induction in the soil is stable, it suffers from a low transmission range and therefore requires a multi-hop network set up to
communicate the information to the surface. This technology is still in early-stage and requires further research to enable efficient IoUGT systems.

Another core element of the X-IoT system is the space network consisting of satellites that can further disseminate sensing data coming from different IoT networks, such as monitoring data on solar activity and changes in the atmosphere, land, oceans, and glaciers, tracking of carbon emissions, and global temperature changes.
Indeed, a substantial number of satellite missions have already been initiated to observe these critical variables \cite{satdata}. Each
of these missions consists of either a single satellite or a constellation of satellites equipped with sensors operating at an infrared/visible range of frequencies.
Recent studies show that the constellation of small satellites with low cost can provide valuable information concerning climate change \cite{Sweeting2018}.
For instance, the cyclone global navigation system  (CYGNSS) developed by the National Aeronautics and Space Administration (NASA) uses small satellites that collect reflected Global Positioning System (GPS) signals, providing information on wind speed over the oceans, flooding, and soil moisture \cite{ruf2018new}. Some other examples include the Dynamic Ionosphere CubeSat Experiment (DICE) mission consisting of small satellites that monitor the Earth’s ionosphere and identifies storms in the geospace. Another small satellite mission called QaukeSat utilized a magnetometer on board to oversee the seismic activities. A list of these missions for climate monitoring can be found in the Nano-satellite database

These satellites use different frequency bands to share the information with the ground station and with the other satellites. These bands start from high frequencies, i.e., 20 MHz, and extend to the optical frequency band. Each of these different bands behaves differently to the absorption, scattering, and precipitation of the atmosphere. Notably, the rain causes high absorption and scattering for the electromagnetic waves that lead to high attenuation of the signal. Rainfall effects are significant till 100 GHz of frequency, after which the attenuation does not increase noticeably. Hence, for countries with a high rain rate, require a suitable selection of frequency to avoid high attenuation. Sky noise, which is mainly caused by sun radiations, is
another main parameter affecting the propagation of signals. Accordingly, the satellite antennas should avoid pointing towards the sun to reduce the sky noise. The frequency band below 100 MHz is usually not considered for the climate monitoring missions due to the cosmic noise, man-made noise, and high ionospheric effects in this band. Atmospheric absorption is low in the
frequency band of 100 MHz to 1 GHz; however, background noise is relatively high. The lower
end of the 1-10 GHz frequency band has little atmospheric effects and therefore are much suited
for the satellite to ground communications. Moving above 10 GHz leads to high attenuation due to gaseous absorption and atmospheric precipitation. More relevant information about the propagation
characteristics of different frequency bands for climate monitoring missions is available
in the ITU-R report \cite{Ospin2011}. Recent studies show that multi-band communication across various
frequency bands allows the use of power and spectrum efficiently; however, this approach is in
the very early stage of the research (6). Besides the use of different frequency bands for communications,
networking among the small satellites is also a demanding task. Currently, communication
among the small satellites is achieved by either using electromagnetic or optical waves.
The former is sensitive to the on-board electronics and cannot support high-speed connections
while the latter requires accurate pointing and acquisition methods among the satellites. One
significant example of satellite-to-satellite communication is QB-50 mission that monitors the
thermosphere of earth.
Based on these space missions, it is crucial to establish a globally connected IoST network
where the sensing information is integrated from the different satellite networks to monitor the
climate. The integration among these missions will be a challenging task since they operate on
a different type of communication technology and have distinct networking architectures.

\section*{\small{Localization}}
The locations where sensing data is collected is another important property for climate monitoring. For example, in the case of measuring the ocean’s salinity, it is critical to know where the
sensors have collected this data. Thanks to the availability of GPS signals, sensing data can usually be geographically tagged. However, in more difficult environments such as underwater and underground, these signals cannot penetrate the surrounding matter and therefore require other means of finding a sensor's location. 
Hence, other methods have been developed to geo-tag sensing data in such difficult environments.  Localization methods in IoUT networks use ranging information calculated from either the time, angle, or power of the acoustic, optical, and magnetic induction signals, which is then used by multilateration and triangulation techniques to estimate the location of underwater IoUT devices \cite{SAEED2019101935}.  
A similar approach is used in IoUGT networks; however, only magnetic induction and acoustic technologies are capable of geo-tagging in IoUGT networks \cite{saeed2019towards}. The estimation of a satellite's location is also critical to efficiently communicate with ground stations and other satellites. Based on the
orbital parameters of the spacecraft, such as altitude, inclination, eccentricity, and velocity, the
location of the satellites can be continuously tracked from the ground \cite{saeed2019cubesat}. Moreover, the data from
the satellites is geo-tagged by the GPS signals, which can significantly help in global climate
monitoring.
\section*{\small{Climate-Friendly IoT Sensors}}
The choice of the type of sensing IoT device needed for radically different environments is critical for climate monitoring. Sensors should be developed from non-toxic, biodegradable materials such that once they reach the end of their lifetime, they are not harmful to the environment and contribute to local pollution. Recent studies have shown that it is possible to create biodegradable electronic circuits and sensors \cite{Dincer2019}. Such a large number of IoT devices require a massive amount of external power to operate, consequently increasing their carbon footprint; ICT technology can use up to 10\% of the global electricity budget \cite{BELKHIR2018448}. Hence, sensors that can extract energy from the environment, i.e., from air pressure, light, or temperature, would help to reduce the sensors' carbon footprint. For instance, the energy harvesting for the optical waves based IoUT network shows that it not only reduces the carbon footprint but also improves the localization accuracy \cite{Saeedtwc}. Nevertheless, the research is still in its infancy on developing micro-systems that are energy autonomous to power the IoT devices.

\section*{\small{Acknowledgments}}
This work is supported by Office of Sponsored Research (OSR) at King Abdullah University of Science and Technology (KAUST). Fig. 1  was produced by Xavier Pita, scientific illustrator at KAUST.

\bibliographystyle{IEEEtran}
\bibliography{scibib}


\end{document}